\documentclass[aps,twocolumn,showpacs]{revtex4}
\usepackage{graphicx}
\usepackage{dcolumn}% Align table columns on decimal point
\usepackage{bm}% bold math
\usepackage{xcolor}

\begin{document}

\title{Exploring chaotic properties in the nonlinear Walecka Model}

\author{Luiz Ant\^onio Barreiro$^{1}$ and  Andr\'e L.\ P.\ Livorati$^{1}$}

\affiliation{$^1$ São Paulo State University (UNESP), Institute of Geosciences and Exact Sciences, Rio Claro, Brazil.
\\
}

\begin{abstract}

This paper investigates dynamical properties of the Nonlinear Walecka Model (NLWM), also known as Quantum Hadrodynamics (QHD), which provides a relativistic framework for describing nuclear matter through nucleon-meson interactions. While the original linear model successfully captures qualitative features of nuclear matter, it overestimates nuclear compressibility; consequently, nonlinear extensions incorporating cubic and quartic self-interaction terms in the scalar field $(\sigma)$ are employed to achieve better agreement with experimental data. By treating the self consistent effective-mass equation as an iterative mapping, the study explores the emergence of complex behaviors such as periodic orbits and chaotic regimes. Using tools from nonlinear dynamics, including return maps and bifurcation diagrams, we characterize routes to chaos via period doubling cascades and identify the presence of "shrimps" stable isoperiodic islands within chaotic regions of the parameter space. Furthermore, we discusses the physical implications of these findings, suggesting that crisis phenomena and structural convergence may serve as dynamical signatures of macroscopic rearrangements in the scalar mean field, potentially influencing the stability and equation of state (EOS) of dense nuclear matter in extreme environments like neutron stars.

\end{abstract}

\maketitle

\section{Introduction}
\label{sec1}

The Walecka Model, often referred to as Quantum Hadrodynamics (QHD), holds a central place in the theoretical study of nuclear matter, offering a relativistic framework to understand the interactions between nucleons (protons and neutrons) mediated by mesons. This model, first introduced by John Dirk Walecka in the mid 1970s, provides a quantum field theoretical description of nuclear forces that accounts for the relativistic effects crucial at high densities, making it an essential tool in both nuclear physics and astrophysics \cite{ref1}.

Prior to the development of the Walecka Model, nuclear matter was typically described using non-relativistic models such as the liquid-drop model and Skyrme forces. These models, while effective in explaining basic nuclear properties, did not account for the underlying relativistic nature of nuclear forces. Walecka's model addressed this by incorporating the exchange of scalar ( $\sigma$ ) mesons, which provide an attractive force between nucleons, and vector $(\omega)$ mesons, which introduce a repulsive force. This interaction, governed by a relativistic Lagrangian, allows for a mean-field treatment that describes nuclear matter in terms of meson fields interacting with nucleons \cite{ref2}.

One can ask about the connection between quantum hadronic models and nonlinear dynamics, and how the effective field equations in nuclear physics (often nonlinear due to many-body interactions) manifest complex behavior? Indeed, nonlinear dynamics are essential in reproducing key features of nuclear matter, such as its binding energy and incompressibility, while also ensuring a more realistic density dependence of the nuclear equation of state (EOS).

One of the significant advancements in the Walecka Model came with the introduction of self-interacting terms, particularly in the interaction between the scalar and vector meson fields. In its original formulation, the Walecka Model (or QHD-I) treated these meson fields in a linear fashion, resulting in certain limitations in its ability to accurately describe the compressibility of nuclear matter and the properties of finite nuclei. This led to the development of QHD-II, where nonlinear self-interactions of the scalar meson field were introduced to improve the model's agreement with experimental data \cite{ref3}. In particular, the addition of cubic and quartic scalar field terms softens the EOS at high densities, preventing the overestimation of the compressibility modulus and improving the description of the nuclear surface \cite{ref4}.

The inclusion of nonlinear scalar self-interactions modifies the model's Lagrangian density by adding terms of the form:
\begin{equation}
\mathcal{L}_{\text {nonlinear }}=-\frac{1}{3} \alpha m_{N} g_{s}^{3} \phi^{3}-\frac{1}{4} \beta g_{s}^{4} \phi^{4}
\label{eq1}
\end{equation}
where $b$ and $c$ are constants, $g_{s}$ is the coupling constant of the scalar meson, $m_{N}$ is the free nucleon mass and $\phi$ is the scalar meson field. These terms introduce important repulsive effects at higher densities, crucial for describing nuclear matter at saturation.

The nonlinear Walecka Model has become especially important in describing extreme environments such as the cores of neutron stars, where densities far exceed normal nuclear matter. At these densities, relativistic effects and nonlinear meson interactions become dominant. By introducing these nonlinear terms, the model provides a more accurate prediction of the equation of state, particularly in balancing the attractive and repulsive forces at high densities \cite{ref5}.

Additionally, nonlinear dynamics allow for the study of phase transitions in dense matter, such as the transition from nuclear matter to hyperonic matter or quark matter. Recent works have used nonlinear extensions of the Walecka Model to explore the possible emergence of new phases in neutron stars, including mixed phases that arise due to the interplay of different forces \cite{ref6}.

The inclusion of nonlinear dynamics also plays a critical role when finite temperature effects are considered. For applications in heavy-ion collisions or astrophysical events such as supernovae, finite temperature formulations of the nonlinear Walecka Model help describe the phase transition from hadronic matter to quark-gluon plasma, which occurs at extremely high temperatures and densities\cite{ref7}. These studies are crucial for understanding the early universe's behavior and the dynamics of matter under extreme conditions.

In recent decades, the Walecka Model has undergone further development, especially in the context of high density astrophysical environments such as neutron stars. Modifications to the original framework, including the addition of new meson fields (such as the isovector $\rho$ meson) and density-dependent coupling constants, have enhanced the model's ability to describe asymmetric nuclear matter, crucial for understanding neutron-rich systems \cite{ref8}. Additionally, studies on the equation of state (EOS) of dense matter, crucial for predicting the mass and radius of neutron stars, have applied the Walecka Model with impressive results \cite{ref9}.

The Walecka Model remains a cornerstone in the study of nuclear matter. Its continued evolution, driven by refinements in observational astrophysics and nuclear experiments, ensures that it remains relevant in the search for new phases of matter under extreme conditions. Future developments are expected to focus on precision treatments of nuclear interactions, the inclusion of higher-order terms, and further exploration of the behavior of exotic nuclear matter.

The transition from orderly nuclear states to chaotic regimes can be described by the emergence of bifurcation sequences in the solutions of effective Lagrangian parameters, where small adjustments in coupling constants lead to fundamental shifts in hadronic stability. Within this framework,
we characterized separate parameter regimes where the presence of shrimp-shaped domains \cite{shrimp1,shrimp2,shrimp3,shrimp4,shrimp5,shrimp6}, reveals that even within the high-energy densities of quantum systems, there exist "islands" of periodic stability. These self-similar structures suggest that the organization between periodic and chaotic behavior is a universal feature that transcends classical systems, offering a geometric perspective on the hierarchical nature of nonlinear interactions in dissipative quantum hadronic models.

In this paper, we undertake a comprehensive analysis of the Walecka Model by exploring its limitations within the framework of nonlinear dynamics. Our study focuses on critical aspects such as periodic and chaotic orbits obtined through return diagrams and their dependence on the inital conditions and the Fermi momemntum $k_F$. We also investigate bifurcation diagrams and routes do chaos by period doubling cascades, where small variations in parameters cause qualitative changes in the behavior of the systems, potentially leading to transitions between different regimes of nuclear matter.
Also, we characterized shrimps shape domains that displays a rich scenario by forming regions of isoperiodic windows. Additionally, we examine the sensitivity of the model to initial conditions, identifying specific configurations that result in divergent or unstable solutions, thereby shedding light on the model's robustness and the boundaries of its applicability in high density and extreme environments. This approach allows us to better understand the nonlinearities inherent in the model and their implications for both theoretical predictions and astrophysical applications.

This paper is organized as follows: Section \ref{sec2} is devoted to display the analytical description of the nonlinear Walecka model, considering Lagrangian approach and the equations of motion. The chaotic properties, the mapping under study, the bifurcation diagrams, shrimp shape domains and other results are displayed in Sec.\ref{sec3}. Finally, in Sec. \ref{sec4} we drawn some discussion, final remarks and conclusions.

\section{ Nonlinear Walecka Model}
\label{sec2}
The original Walecka model is purely linear and, while successful in capturing the qualitative aspects of nuclear matter, fails to match quantitative results such as the compressibility of nuclear matter. To address these shortcomings, the Non-Linear Walecka Model (NLWM) \cite{ref3} was developed, introducing self-interaction terms for the scalar meson field. This extension improves the model's predictions for nuclear saturation properties and compressibility, providing a more accurate description of dense nuclear matter.

\subsection{Lagrangian density}
The NLWM describes the dynamics of nucleons (represented by the Dirac field $\psi$ ) and their interactions through scalar ( $\sigma$ ) and vector ( $\omega_{\mu}$ ) meson fields. The Lagrangian density includes nonlinear self-interaction terms for the scalar meson field and is expressed as follows:
\begin{equation}
\begin{array}{lll}
\mathcal{L}= & \bar{\psi}\left(i \gamma^{\mu} \partial_{\mu}-m_{N}-g_{\sigma} \sigma-g_{\omega} \gamma^{\mu} \omega_{\mu}\right) \psi \\
& +\frac{1}{2}\left(\partial_{\mu} \sigma\right)\left(\partial^{\mu} \sigma\right)-\frac{1}{2} m_{\sigma}^{2} \sigma^{2}-\frac{\bar{\alpha}}{3} m_{N} g_{\sigma}^{3} \sigma^{3}-\frac{\bar{\beta}}{4} g_{\sigma}^{4} \sigma^{4} \\
& -\frac{1}{4} F_{\mu \nu} F^{\mu \nu}+\frac{1}{2} m_{\omega}^{2} \omega_{\mu} \omega^{\mu}
\end{array}
\label{eq2}
\end{equation}
where the Einstein summation convention is applied, such that repeated indices (covariant and contra variant) are summed over the spacetime coordinates. Here $\psi$ is the nucleon field, $\sigma$ is the scalar meson field (associated with the attractive nuclear force), $\omega_{\mu}$ is the vector meson field (associated with the repulsive nuclear force), $g_{\sigma}$ and $g_{\omega}$ are the coupling constants for the scalar and vector mesons, respectively, $m_{N}, m_{\sigma}$ and $m_{\omega}$ are the respective masses of the nucleons, sigma and omega mesons, $F_{\mu \nu}=\partial_{\mu} \omega_{\nu}-\partial_{\nu} \omega_{\mu}$ is the field tensor for the vector meson and $\bar{\alpha}$ and $\bar{\beta}$ are parameters that describe the strength of the non-linear self-interactions of the sigma meson.

The nucleons interact with the scalar ( $\sigma$ ) and vector $\left(\omega_{\mu}\right)$ mesons. The scalar field ( $\sigma$ ) induces an attractive interaction, which is necessary for binding nucleons together in nuclei, while the vector field ( $\omega_{\mu}$ ) generates a short range repulsive interaction that becomes significant at high densities. The inclusion of cubic $\left(g_{\sigma}^{3} \sigma^{3}\right)$ and quartic $\left(g_{\sigma}^{4} \sigma^{4}\right)$ terms introduces self-interactions for the scalar field. These terms are crucial in reducing the compressibility of nuclear matter to realistic values, making the model more accurate for dense systems. The $m_{\sigma}^{2} \sigma^{2}$ and $m_{\omega}^{2} \omega_{\mu} \omega^{\mu}$ terms represent the mass terms for the scalar and vector mesons, respectively. These terms are critical in determining the range of the interactions mediated by these mesons. The field strength tensor $F_{\mu \nu}$ describes the dynamics of the vector meson field, ensuring that the model respects the gauge symmetry associated with the vector meson.

\subsection{Equations of motion}
The evolution of a physical system follows the principle that the action is minimized (or rendered stationary) with respect to infinitesimal variations in the system's configuration. This leads to the derivation of the Euler Lagrange equations, which, for a general field, are expressed as
\begin{equation}
\frac{\partial \mathcal{L}}{\partial \phi_{i}}-\partial_{\mu}\left(\frac{\partial \mathcal{L}}{\partial\left(\partial_{\mu} \phi_{i}\right)}\right)=0 .
\label{eq2.1}
\end{equation}
The index $i$ denotes the components of the field. For a scalar field, there is a single component, $\phi_{i}=\sigma$. In the case of a vector field, $i$ represents the four space-time components, $\phi_{i}=\omega_{\mu}$, while for a spinor field, the index specifies the particle $\psi$ or antiparticle $\bar{\psi}$. Consequently, by applying the Euler-Lagrange equation to the Walecka model, we derive the following equations of motion:

1. Using $\phi_{i}=\bar{\psi}$ we obtain the Nucleon (Dirac) Equation:
\begin{equation}
\left(i \gamma^{\mu} \partial_{\mu}-g_{\omega} \gamma^{\mu} \omega_{\mu}-M^{*}\right) \psi=0
\label{eq3}
\end{equation}
where
\begin{equation}
M^{*}=m_{N}+g_{\sigma} \sigma
\label{eq4}
\end{equation}
which establishes the nucleon mass in the medium.\\

2. Using $\phi_{i}=\sigma$ we obtain the Scalar Meson (Klein Gordon) Equation:
\begin{equation}
\partial_{\mu} \partial^{\mu} \sigma+m_{\sigma}^{2} \sigma+\bar{\alpha} m_{N} g_{\sigma}^{3} \sigma^{2}+\bar{\beta} g_{\sigma}^{4} \sigma^{3}=-g_{\sigma} \bar{\psi} \psi
\label{eq5}
\end{equation}

Using $\phi_{i}=\omega_{\mu}$ we obtain the Vector Meson (Proca) Equation:

\begin{equation}
\partial_{\nu} F^{\nu \mu}+m_{\omega}^{2} \omega^{\mu}=g_{\omega} \bar{\psi} \gamma^{\mu} \psi
\label{eq6}
\end{equation}

These equations describe the interactions between nucleons and mesons in the non-linear Walecka model, with nucleon fields sourcing the meson fields and meson fields affecting the motion of the nucleons. This system comprises coupled differential equations that are inherently challenging to solve. However, due to the presence of fermions, the energy levels are populated up to the established Fermi level, denoted by $k_{F}$. As a result, nucleons within the nuclear medium, while interacting, remain in their respective levels, as all other states are occupied. This confinement leads nucleons to exist within a consistent mean field produced by the surrounding nucleons. This approach constitutes the mean field approximation, which is implemented mathematically by substituting the mesonic fields with their expected (mean) values: $\sigma \rightarrow\langle\sigma\rangle=\sigma_{0}$ and $\omega_{\mu} \rightarrow\left\langle\omega_{\mu}\right\rangle=\delta_{\mu 0} \omega_{0}$. For a static and uniform system, the expectation values, denoted as $\sigma_{0}$ and $\omega_{0}$, remain constant. Consequently, equations (5) and (6) can be solved straightforwardly, yielding
\begin{equation}
\begin{array}{ll}
 \sigma_{0} & =-\frac{g_{\sigma}}{m_{\sigma}^{2}}\langle\bar{\psi} \psi\rangle-\bar{\alpha} \frac{g_{\sigma}^{3}}{m_{\sigma}^{2}} \sigma_{0}^{2}-\bar{\beta} \frac{g_{\sigma}^{4}}{m_{\sigma}^{2}} \sigma_{0}^{3} \\
\omega_{0} & =\frac{g_{\omega}}{m_{\omega}^{2}}\left\langle\bar{\psi} \gamma^{0} \psi\right\rangle=\frac{g_{\omega}}{m_{\omega}^{2}}\left\langle\psi^{\dagger} \psi\right\rangle
\end{array}
\label{eq7}
\end{equation}
where $\langle \bar{\psi}\psi \rangle$ and $\left\langle\psi^{\dagger} \psi\right\rangle$ denote the scalar density and the baryon (vector) density, respectively.

The expectation values in the ground-state of symmetric nuclear matter are given by \cite{ref1,ref2,ref3}
\begin{eqnarray}
\left\langle \psi^{\dagger}\psi \right\rangle
&=&
\frac{2}{3\pi^{2}}\,k_{F}^{3},
\label{Barionic}
\\[6pt]
\left\langle \bar{\psi}\psi \right\rangle
&=&
\frac{M^{*}}{\pi^{2}}
\left[
k_{F}E_{F}^{*}
-
{M^{*}}^{2}
\ln\!\left(
\frac{k_{F}+E_{F}^{*}}{M^{*}}
\right)
\right],
\label{Scalar}
\end{eqnarray}
where the effective Fermi energy is defined as $E_{F}^{*}=\sqrt{k_{F}^{2}+{M^{*}}^{2}}$.

Using these expressions, the self-consistent equation for the in-medium nucleon mass becomes
\begin{equation}
\begin{array}{ll}
M^{*}= & m_{N}-\frac{g_{\sigma}^{2}}{m_{\sigma}^{2}}\left\{\frac{M^{*}}{\pi^{2}}\left[k_{F} E_{F}^{*}-M^{*^{2}} \ln \left(\frac{k_{F}+E_{F}^{*}}{M^{*}}\right)\right]\right. \\
& \left.+\bar{\alpha} m_{N} g_{\sigma}^{2}\left(M^{*}-m_{N}\right)^{2}+\bar{\beta} g_{\sigma}^{3}\left(M^{*}-m_{N}\right)^{3}\right\}
\end{array}
\label{eq9}
\end{equation}

Equation~(\ref{eq9}) must be solved numerically in a self-consistent manner. Starting from an initial guess for $M^{*}$, an iterative procedure is applied until convergence to a stable fixed point is achieved. This formulation naturally allows the use of nonlinear dynamical tools to analyze the structure of solutions, including the identification of attraction basins, bifurcation points, and parameter thresholds.

The energy density and pressure of the nonlinear Walecka model
are obtained from the canonical (Noether) energy–momentum tensor
\begin{equation}
T^{\mu\nu}
=
\sum_i
\frac{\partial \mathcal{L}}{\partial (\partial_\mu \phi_i)}
\, \partial^\nu \phi_i
-
g^{\mu\nu}\mathcal{L}.
\end{equation}

In uniform and static nuclear matter (mean-field approximation),
the energy density is given by $\varepsilon = T^{00}$, yielding
\begin{equation}
\varepsilon
=
\varepsilon_N(M^*)
+
\frac{1}{2} m_\sigma^2 \sigma_0^2
+
\frac{\bar{\alpha}}{3} m_N g_\sigma^3 \sigma_0^3
+
\frac{\bar{\beta}}{4} g_\sigma^4 \sigma_0^4
+
\frac{1}{2} m_\omega^2 \omega_0^2 ,
\end{equation}
where the nucleonic contribution reads
\begin{equation}
\varepsilon_N(M^*)
=
\frac{k_F E_F^*}{4\pi^{2}} (2k_F^2 + {M^*}^2)
-
\frac{ {M^*}^4}{4\pi^{2}}
\ln\!\left(
\frac{k_F + E_F^*}{M^*}
\right).
\end{equation}

The pressure is obtained from
\(
P = \frac{1}{3} T^{ii}
\)
and can be written as
\begin{equation}
P
=
P_N(M^*)
-
\frac{1}{2} m_\sigma^2 \sigma_0^2
-
\frac{\bar{\alpha}}{3} m_N g_\sigma^3 \sigma_0^3
-
\frac{\bar{\beta}}{4} g_\sigma^4 \sigma_0^4
+
\frac{1}{2} m_\omega^2 \omega_0^2 , \label{eqP}
\end{equation}
where the nucleonic contribution is
\begin{equation}
P_N(M^*)
=
\frac{k_F E_F^*}{12\pi^{2}} (2k_F^2 - 3{M^*}^2)
+
\frac{ {M^*}^4}{4\pi^{2}}
\ln\!\left(
\frac{k_F + E_F^*}{M^*}
\right).
\end{equation}

\subsection{Fixed-point iterative formulation of the effective-mass equation}

The transition from the original covariant field equations, which are complicated coupled partial differential equations (PDEs), to a discrete one-dimensional map is made through the Relativistic Mean-Field (RMF) approximation. Under the assumption of a static, uniform, and infinite nuclear medium, the space-time derivatives of the meson fields vanish ($\partial_\mu \sigma \rightarrow 0, \partial_\mu \omega_\nu \rightarrow 0$), reducing the differential equations of motion to static algebraic relations.

The self-consistent effective-mass ($M^*$) equation obtained in Eq.~(\ref{eq9}) is a nonlinear transcendental equation owing to its logarithmic dependence and nonlinear scalar self-interaction terms. Since an analytical closed-form solution is unavailable, it must be solved numerically via a fixed-point iterative scheme.

To this end, Eq.~(\ref{eq9}) can be rewritten in the generic fixed-point form
\begin{equation}
M^{*}=f(M^{*};k_{F},\alpha,\beta,G),
\end{equation}
where $f$ represents the nonlinear functional defined by the right-hand side of Eq.~(\ref{eq9}). The standard numerical implementation of the self-consistency condition is then naturally expressed as the iterative map
\begin{equation}
M_{n+1}=f(M_{n}),
\label{iteration}
\end{equation}
where the integer index $n$ labels successive iterations of the numerical self-consistency cycle.

It is important to emphasize that the discrete index $n$ does not represent the physical time evolution of nuclear matter or nucleon fields. Instead, it corresponds exclusively to the sequence of successive iterations within the numerical self-consistency cycle. In this sense, the mapping introduced in Eq.~(\ref{iteration}) is not an additional phenomenological assumption, but rather the direct numerical realization of the mean-field problem itself.

Iterative fixed-point schemes of this type are ubiquitous in many-body physics and quantum field theory, appearing in Hartree--Fock calculations \cite{HFC0,HFC1}, density-functional theory \cite{DFT}, BCS (Bardeen-Cooper-Schrieffer) gap equations \cite{BCS1,BCS2}, Dyson--Schwinger equations \cite{DSE1,DSE2}, and relativistic mean-field models \cite{ref2}. In such approaches, the convergence properties of the iterative procedure are closely related to the structure and stability of the underlying nonlinear equations.

Within this framework, a stable fixed point of Eq.~(\ref{iteration}) corresponds to a convergent self-consistent solution for the effective nucleon mass. Conversely, loss of stability of a fixed point may signal the emergence of competing self-consistent branches or sensitivity to initial conditions in the iterative process. Therefore, the dynamical-systems analysis performed in this work should be interpreted as an investigation of the nonlinear structure of the self-consistent solution space of the nonlinear Walecka model, rather than as a microscopic dynamical evolution law for nuclear matter.

Accordingly, the bifurcation diagrams, return maps, and stability analyses presented in the following sections provide information about how the set of admissible self-consistent solutions reorganizes as the control parameters $(k_{F},\alpha,\beta)$ are varied. In particular, the appearance of periodic attracting fixed points elements in the iterative mapping space, reflects the coexistence of distinct self-consistent branches of the scalar mean field.

\section{Nonlinear Analysis of the Self-Consistent Map}
\label{sec3}

\subsection{The iterative map}

As discussed in the previous section, the self-consistent effective-mass equation of the nonlinear Walecka model cannot, in general, be solved analytically. The determination of the in-medium nucleon mass therefore requires the implementation of a fixed-point iterative procedure. Rewriting Eq.~(\ref{eq9}) in iterative form leads naturally to the mapping
\begin{equation}
M_{n+1}=f\left(M_{n}\right),
\label{eq10}
\end{equation}
where
\begin{equation}
\begin{array}{lll}
f\left(M_{n}\right)= & m_{N} 
 -\left\{\bar{G} M_{n}\left[k_{F} E_{n}-M_{n}^{2} \ln \left(\frac{k_{F}+E_{n}}{M_{n}}\right)\right]\right. \\[0.2cm]
& \left.+{\alpha} m_{N}\left(M_{n}-m_{N}\right)^{2}+{\beta}\left(M_{n}-m_{N}\right)^{3}\right\},
\end{array}
\label{eq11}
\end{equation}
with
\begin{equation}
E_{n}=\sqrt{k_{F}^{2}+M_{n}^{2}},
\end{equation}
and where the parameters were redefined as
\begin{equation}
{G}=\frac{g_{\sigma}^{2}}{m_{\sigma}^{2}\pi^{2}},
\qquad
\alpha=\bar{\alpha}\frac{g_{\sigma}^{4}}{m_{\sigma}^{2}},
\qquad
\beta=\bar{\beta}\frac{g_{\sigma}^{5}}{m_{\sigma}^{2}}.
\end{equation}

The integer index $n$ labels successive iterations of the numerical self-consistency cycle and does not correspond to the physical time evolution of nuclear matter. In this framework, convergence of the sequence $\{M_n\}$ toward a stable fixed point corresponds to the existence of a self-consistent mean-field solution for the effective nucleon mass.

The purpose of the present analysis is therefore to investigate the nonlinear structure and stability properties of the self-consistent iterative map associated with the nonlinear Walecka model. In particular, we examine how the organization of fixed points, periodic orbits, and stable regions changes under variations of the control parameters $(k_F,\alpha,\beta)$.

It is worth emphasizing that the dynamical structure explored here is qualitatively distinct from the chaotic behavior found in generic numerical root-finding schemes, such as the Newton–Raphson method or iterative searches for the roots of a fixed polynomial equation. In those cases, the fractal or chaotic structure is a consequence of the choice of algorithm and of the initial guess for a fixed equation, and carries no dependence on any physical parameter of the problem; a different iterative scheme applied to the same equation would, in general, yield an entirely different convergence structure without altering the underlying physics.

In the present analysis of our study, by contrast, the effective-mass equation possesses  attracting fixed points, bifurcations, periodic windows, and chaotic regimes for some values of physical configuration $(k_F, \alpha, \beta)$, as shown in the next sections. As for example, one could apply the cobweb convergence in Fig.\ref{fig1}(d), independent of the initial guess $M_0$.

This distinction is consistent with previous studies that interpret self-consistent-field iterations of physical origin as genuine dynamical systems whose parameter dependence reflects the underlying physics rather than numerical artifacts of the solution algorithm \cite{claude1,claude2}. It is further consistent with the demonstration, within the Walecka model itself, that multiple solutions of the self-consistency condition correspond to a genuine first-order phase transition of nuclear matter rather than to numerical degeneracy \cite{claude3}.

Due to the strong nonlinearity introduced by the logarithmic contribution and by the cubic and quartic scalar self-interaction terms, the iterative map may undergo loss of fixed-point stability as the parameters are varied. Such transitions can generate multiple self-consistent branches, bifurcation structures, and complex convergence patterns within the iterative scheme. The present work extends this line of investigation by systematically mapping, through bifurcation diagrams, Lyapunov exponents, and isoperiodic (shrimp-shaped) domains, the full parameter-space organization of this nonlinear solution structure. Consequently, tools from nonlinear dynamics provide a convenient mathematical framework to characterize the stability and organization of the self-consistent solutions of the model.

\subsection{Fixed Points and Bifurcations}

A fixed point (or period-1 orbit) is a value $M^{*}$ such that, when the map of the system is applied, the point maps into itself:
\[
M^{*} = f\!\left(M^{*}\right).
\]
This condition implies that if the iteration starts at $M^{*}$, it remains there for all subsequent iterations, characterizing a period-1 orbit \cite{meiss,litch,hilborn,chaos}.

\begin{figure}[h]
\begin{center}
\centerline{\includegraphics[width=80mm,height=40mm]{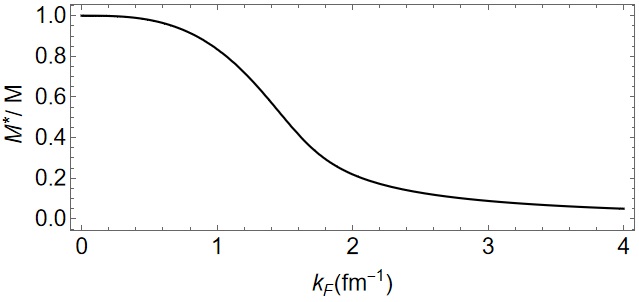}}
\end{center}
\caption{ {\it Effective nucleon mass in symmetric nuclear matter, $M^*/M$, normalized by the vacuum bare nucleon mass $M \approx 939 \text{ MeV}$, as a function of the Fermi momentum $k_F$ in the Walecka (relativistic mean-field) model with $\alpha=\beta=0$. The decrease of $M^{*}$ with increasing
$k_F$ reflects the growth of the attractive scalar mean field
($\sigma$-meson exchange), which reduces the Dirac mass according to
Eq. (\ref{eq4}). In the low-density limit
($k_F \to 0$), $M^{*} \to M$.}}
\label{fig0}
\end{figure}

In Fig.\ref{fig0}, we first consider the case $\alpha=\beta=0$ and $G=3 \times 10^{-5}\,\mathrm{MeV}^{-2}$.
the effective nucleon mass $M^*$ is normalized by $M$, where $M \approx 939 \text{ MeV}$ represents the bare nucleon mass in vacuum. The ratio $M^*/M$ is the standard dimensionless effective mass parameter widely used in nuclear physics to characterize the strength of the scalar in-medium interactions.
In this regime, the effective mass $M^{*}$ becomes a function of the Fermi momentum $k_F$.
Physically, this corresponds to the standard relativistic mean-field solution of the Walecka (quantum hadrodynamics) model.

The monotonic decrease of $M^{*}/M$ with increasing $k_F$ is a direct consequence of the strengthening of the attractive scalar mean field generated by $\sigma$-meson exchange. In the relativistic mean-field approximation, the effective mass satisfies Eq.~(\ref{eq4}), with the scalar field replaced by its expectation value given by Eq. (\ref{eq7}). Since $\sigma_0$ is determined self-consistently by the scalar density, it increases in magnitude as the baryon density (or equivalently $k_F$) increases. This amplifies the attractive scalar interaction and leads to a substantial reduction of the in-medium nucleon mass.

In the low-density limit ($k_F \to 0$), the scalar density vanishes, implying $\sigma_0 \to 0$ and therefore $M^{*} \to M$, recovering the vacuum nucleon mass. Near nuclear saturation density (corresponding to $k_F \sim 1.3\text{--}1.4\,\mathrm{fm}^{-1}$), the effective mass is already significantly reduced, typically $M^{*}/M \sim 0.6\text{--}0.7$. This reduction is a key ingredient of the relativistic saturation mechanism and strongly influences the structure of the nuclear equation of state. At higher densities, the continued decrease of $M^{*}$ signals the increasing importance of scalar correlations and relativistic effects in dense nuclear matter.

\begin{figure}[ht!]
\begin{center}
\centerline{\includegraphics[width=9cm,height=15cm]{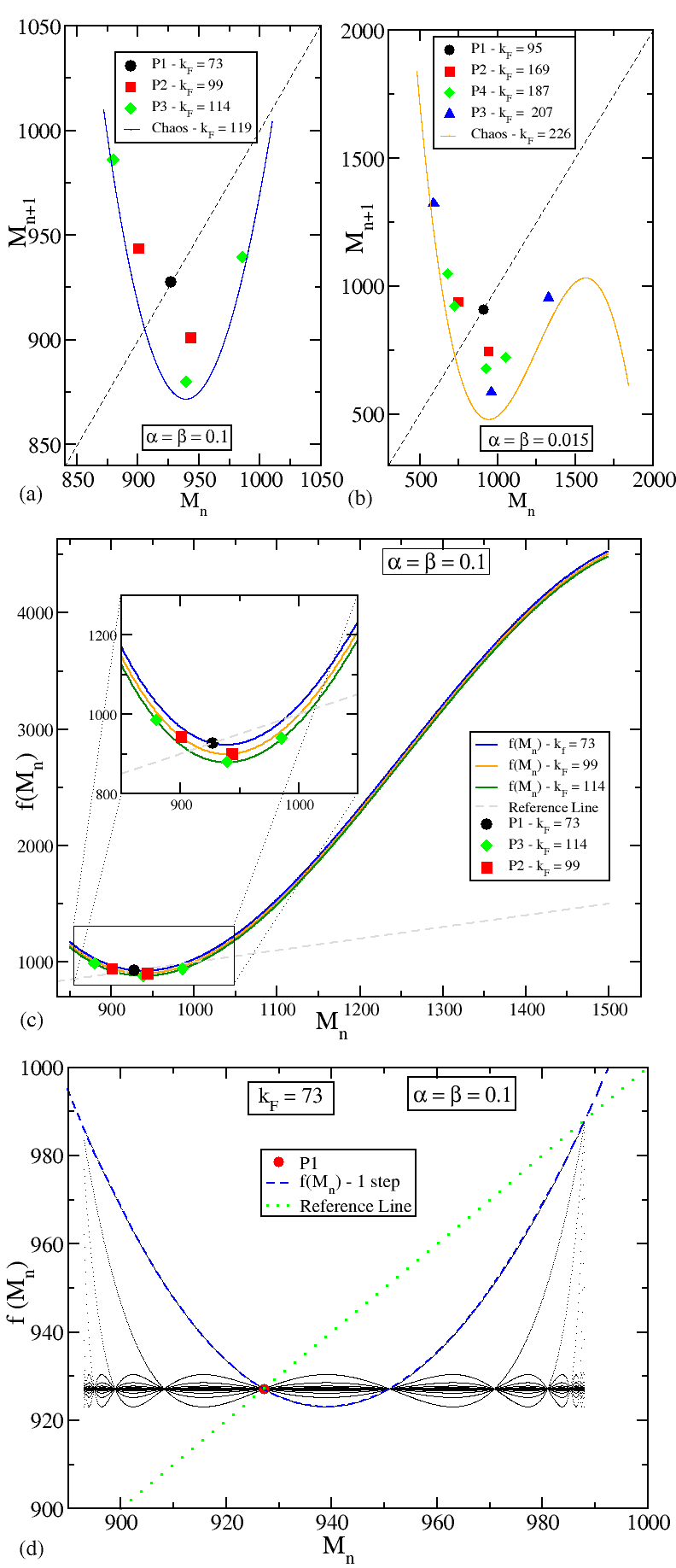}}
\end{center}
\caption{Color online: {\it Iterative mapping structure and fixed-point dynamics in the $(M_n, M_{n+1})$ plane. Panels (a) and (b) display the discrete periodic orbits ($P_1, P_2, P_3, P_4$) and chaotic sets relative to the identity reference line ($M_{n+1} = M_n$) for coupling parameters $\alpha = \beta = 0.1$ and $\alpha = \beta = 0.015$, respectively. Panel (c) shows the continuous one-step mapping functions $f(M_n)$ for $\alpha = \beta = 0.1$ at representative Fermi momenta ($k_F = 73, 99, 114\text{ fm}^{-1}$), where the inset highlights the geometric intersections with the identity line that define the fixed points. Panel (d) illustrates the iterative convergence landscape for $k_F = 73\text{ fm}^{-1}$ over 20 steps across multiple initial conditions $M_0$, demonstrating the symmetrical node contraction toward the period-1 fixed point $P_1$ (red circle). Also, one could apply the cobweb method between the one-step mapping $f(M_n)$ (blue curve)  and the reference line (dashed green) and check the convergence for the $P_1$ attracting fixed point. In all panels, $M_n$ is set to be measured in $MeV$.}}
\label{fig1}
\end{figure}

From the dynamical systems perspective adopted here, the curve in Fig.~\ref{fig0} therefore represents the branch of fixed points of the iterative map defining the self-consistent effective mass. Subsequent bifurcations correspond to the emergence of additional solutions of the fixed-point equation as the control parameters are varied.

When the nonlinear couplings $\alpha$ and $\beta$ are switched on, the scalar sector of the theory acquires cubic and quartic self-interaction terms. Physically, these terms represent many-body correlations beyond the linear $\sigma$-exchange mechanism of the original Walecka model. The cubic term (controlled by $\alpha$) introduces an asymmetry in the scalar potential, effectively modifying the curvature of the potential around its minimum and allowing additional flexibility in reproducing the empirical saturation properties of nuclear matter. The quartic term (controlled by $\beta$) stabilizes the scalar potential at large field amplitudes, preventing an excessive decrease of $M^{*}$ at high densities and softening the equation of state.

From a nuclear matter perspective, $\alpha$ and $\beta$ regulate the density dependence of the scalar mean field and therefore control the compressibility, the saturation mechanism, and the high-density behavior of the equation of state. In the dynamical systems interpretation adopted here, these nonlinear terms modify the structure of the fixed-point equation for $M^{*}$, potentially generating multiple solutions for a given $k_F$. Consequently, $\alpha$ and $\beta$ act as control parameters that can induce bifurcations in the effective-mass map, leading to the coexistence of different self-consistent branches and richer nonlinear behavior.

In order to illustrate how the dynamics evolves concerning the fixed points, we set up a return diagram \cite{meiss,litch,hilborn,chaos}, i.e., a plot of $M_n \times M_{n+1}$, according Eqs.(\ref{eq10}) and (\ref{eq11}), along with the identity bisector line $M_{n+1} = M_n$. The intersections between the map function and this diagonal line directly verify the exact locations of the fixed points, considering a few combinations of the control parameters. Now, instead of considering $\alpha=\beta=0$, we are ranging these parameters and kepping $G=3 \times 10^{-5} \mathrm{MeV}^{-2}$ as a constant. Figure \ref{fig1}(a,b) was constructed considering a grid of $10000$ initial conditions distributed along $M_n=(939 \pm \Delta)~MeV$, where $\Delta=100$, evolved up to a transient of $10^4$ iterations of Eq.(\ref{eq11}), where we marked the final $10^3$ points.

One can see in Fig.\ref{fig1}(a) that
for a range of $k_F$ between $k_F=73~\mathrm{fm}^{-1}$ and $k_F=119~\mathrm{fm}^{-1}$, for $\alpha=\beta=0.1$, the return diagram furnishes us fixed points of different periods, and also an orbit with no determied period (larger then $100$), that we set as chaotic orbit. The same applies to Fig.\ref{fig1}(b) where the range of $k_F$ is now between $k_F=95 ~\mathrm{fm}^{-1}$ and $k_F=226~\mathrm{fm}^{-1}$,
for $\alpha=\beta=0.015$. One can realize that the size of the chaotic orbit, set as an undetermined period orbit (continuous line) increases as with $\alpha=\beta$ diminishes, and it seems that the dynamics are significantly changing as we range $k_F$.

To explicitly illustrate the geometrical origin of the fixed points and their iterative convergence properties, Figs.\ref{fig1}(c) and \ref{fig1}(d) detail the functional form and dynamics of the one-step mapping $f(M_n)$. In Fig.\ref{fig1}(c), the continuous mapping function $f(M_n)$ is plotted for $\alpha = \beta = 0.1$ across representative Fermi momenta ($k_F = 73, 99$, and $114~\text{fm}^{-1}$), with the inset offering a magnified view near the local minimum. This panel explicitly shows that the period-1 fixed point $P_1$ corresponds to the direct intersection between the curve $f(M_n)$ and the diagonal reference line ($M_{n+1} = M_n$), while the elements of higher-order periodic orbits ($P_2, P_3$) lie along their respective functional curves.

\begin{figure*}[htb]
\begin{center}
\centerline{\includegraphics[width=18cm,height=10cm]{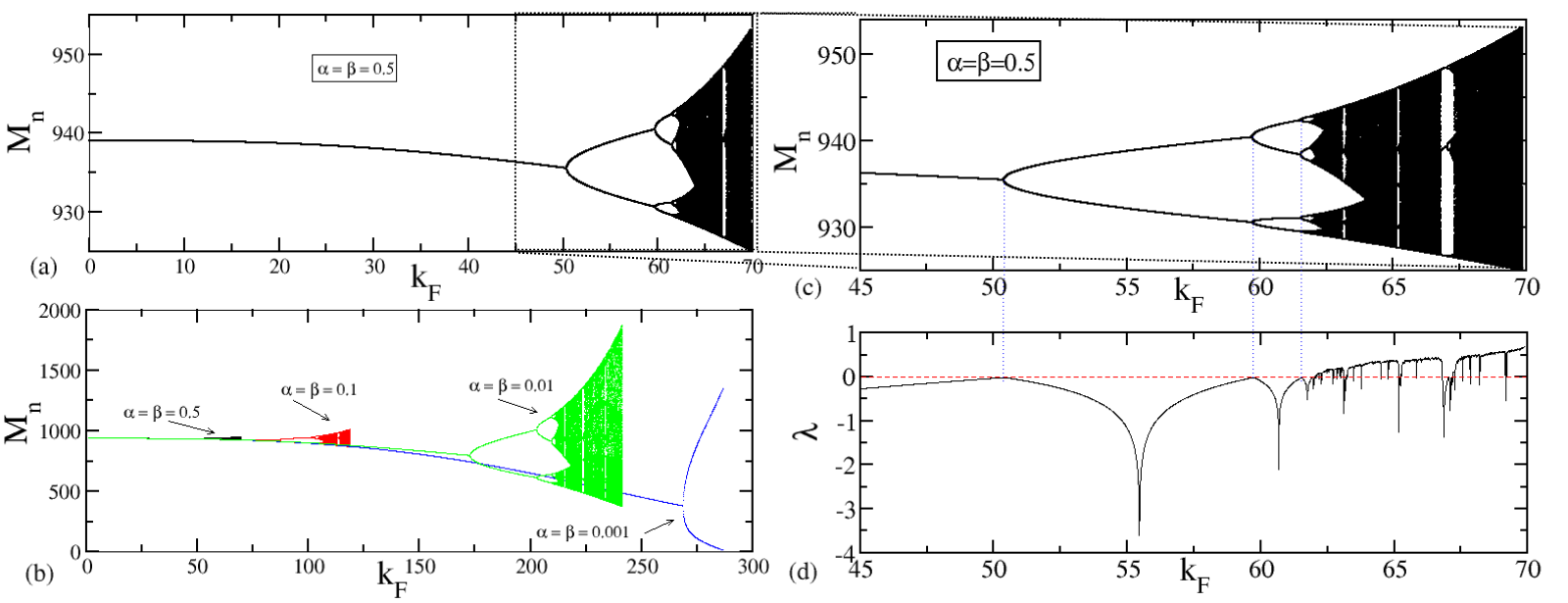}}
\end{center}
\caption{Color online: {\it Route to chaos and bifurcation diagrams as function of the Fermi momentum $k_F \mathrm{fm}^{-1}$. In (a) we have a typical period doubling route to chaos for $\alpha=\beta=0.5$, and in (b) a few bifurcation diagrams for other combination of control parameters. One can see that as $\alpha=\beta$ are getting smaller, the size ($M_n$ axis) and length ($k_F$ axis) are increased. Also, in (b) all diagrams follows the same pattern, and vanishes in a specific parameter of $k_F$. In both, (a) and (b), $k_F$ is measured in $\mathrm{fm}^{-1}$}, and $M_n$ is given in $MeV$.}
\label{fig2}
\end{figure*}

Complementing this, Fig.\ref{fig1}(d) depicts the iterative trajectory landscape for $k_F = 73~\text{fm}^{-1}$ across $20$ mapping steps (in black) for a broad spectrum of initial conditions $M_0$, while the one-step mapping $f(M_n)$ is given in blue. One could apply the cobweb mapping \cite{hilborn,chaos} between the blue curve and the reference line (in green), and check the convergence for the $P_1\approx927.1425.$
The progressive folding of the composed curves $f^n(M_0)$ (set as black curves) toward the intersection coordinate $P_1$ confirms that the period-1 state acts as an attracting fixed point, ensuring that any physically admissible initial trial mass monotonically converges to the unique mean-field equilibrium.

On the other hand, coming back to Fig.\ref{fig1}(a,b) the finding of an orbit with undetermined period in the return diagram, indicates that our system can be chaotic for some combinations of the control pameters. In chaotic systems, fixed points often become unstable as parameters change \cite{bifurca0,bifurca1,bifurca2}, serving as a precursor to more complex behavior \cite{meiss,litch,hilborn,chaos}. So, seeking for a better understanding and visualization of the dynamics thresholds, we contructed some bifurcation diagrams for the Fermi momentum $k_F$, as we ranged $\alpha$ and $\beta$, in search of identifying transitions to chaos.

To check
whether the orbits shown in Fig.\ref{fig1} are chaotic, the Lyapunov
exponent must be evaluated. It is known that the Lyapunov
exponent quantifies the average exponential rate of divergence or convergence of infinitesimally close trajectories. A positive Lyapunov exponent ($\lambda > 0$) signifies a sensitive dependence on initial conditions and serves as a primary diagnostic indicator of deterministic chaos. Conversely, $\lambda < 0$ implies that the system converges toward a stable fixed point or periodic orbit, while $\lambda = 0$ corresponds to a bifurcation point. It is formally defined as:

\begin{equation}
 \lambda = \lim_{N \to \infty} \frac{1}{N} \sum_{n=0}^{N-1} \ln |f'(M_n)|~,
 \label{eq_lyap}
\end{equation}

where $N$ is number of iterations and $f'(M_n)$ is the derivative of Eq.(\ref{eq11}), and is set as

\begin{equation}
\renewcommand{\arraystretch}{1.5}
\begin{array}{l}
f'(M_n) = -G \left[ k_F E_n + \frac{2 k_F M_n^2}{E_n} - 3 M_n^2 \ln \left( \frac{k_F + E_n}{M_n} \right) \right] \\
\qquad \qquad + 2\alpha m_N (m_N - M_n) - 3\beta (m_N - M_n)^2,
\end{array}
\label{eq_lyap_deriv}
\end{equation}

where $E_n = \sqrt{k_F^2 + M_n^2}$.

Figure \ref{fig2} displays the complex behaviour of the bicurcation diagrams. Each diagram in both Figs.\ref{fig2}(a,b) was contructed considering a sigle initial condition $M_0=939~MeV$ evolved up to a $10^4$ transient, and than marked additional $10^3$ iterations. The initial condition was kept fixed for all Fermi momenta $k_F$ range. One can see in Fig.\ref{fig2}(a) a typical behavior of period doubling route to chaos \cite{meiss,litch,hilborn,chaos} for $\alpha=\beta=0.5$. The attractor exists until $k_F=70~\mathrm{fm}^{-1}$, and then it suddenly vanishes.

\begin{figure}[h!]
\begin{center}
\centerline{\includegraphics[width=9cm,height=8cm]{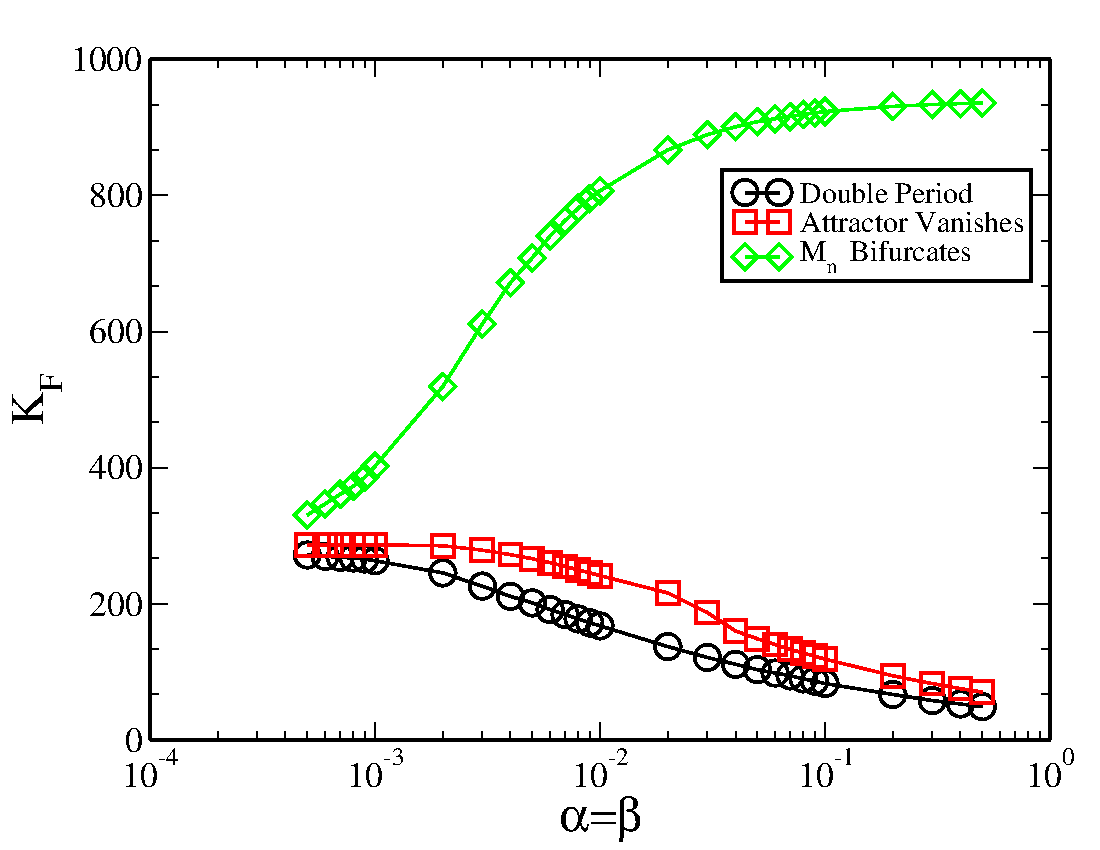}}
\end{center}
\caption{Color online: {\it Characteristic critical values of the period doubling route to chaos for some combinations of the control parameters. The region comprehended between the bullets (where the first bifurcation occurs) and squares (the vanishing of the attractors), is of great interest, since is where the most periodic and chaotic orbits are found. Also, we depicted in diamonds, the value of $M_n$ where the first bifurcation occurs. The vertical axis $k_F$ is measured in $\mathrm{fm}^{-1}$, and the absolute values are the same for the $M_n$ (green diamonds) also in the vertical axis, but measured in $MeV$.}}
\label{fig4}
\end{figure}

A similar behavior is shown in Fig.\ref{fig2}(b), where we now range the values of $\alpha=\beta$ through at least three orders of magnitude to contructed the bifurcation diagrams. One can realize that, when $\alpha=\beta$ diminishes, the size (in the $M_n$ axis) and the length (range of $k_F$)
of the bifurcation diagrams increases. One may compare the distinguisehd periods obtained in Fig.\ref{fig1}, with the results of Fig.\ref{fig2}, and they present a very good agreement, where every periodic orbit and chaotic orbit as well of Fig.\ref{fig1}, can be pointed out in the bifurcation diagram of Fig.\ref{fig2}.

In addition, Fig.\ref{fig2}(c) shows a zoom-in window in the bifurcation diagram of the Fig.\ref{fig2}(a), and a comparison with the Lyapunov exponent, computed by Eqs.(\ref{eq_lyap}) and (\ref{eq_lyap_deriv}) is shown in Fig.\ref{fig2}(d). One can see that the value of $\lambda$ matches according the theory for regular, chaotic and bifurcation points. These results gives robustness to the periodic and chaotic behaviour observed in Figs.\ref{fig1} and \ref{fig2}.

One can also realize that all the bifurcation diagrams of Fig.\ref{fig2}(b), follows the same vanishing pattern, as the control parameter $K_F$ crosses a critical threshold. This sudden destruction of the chaotic attractor is formally identified as a boundary crisis, a well characterized phenomenon in nonlinear dynamics \cite{crisis1,crisis2,crisis3}. This crisis occurs when the chaotic attractor collides with the stable manifold of a coexisting unstable periodic orbit, causing the iterative trajectory to instantly escape the chaotic domain.

A deeper physical insight into the crisis phenomena can be obtained by examining their connection with thermodynamic stability.  Mechanical stability of nuclear matter at zero temperature requires
\[
\left(\frac{\partial P}{\partial \rho_B}\right)_{T=0} > 0,
\]
where $\rho_B=\langle \psi^{\dagger}\psi\rangle$ is the baryon density and $P$ is the pressure density given in Eq. (\ref{eqP}).
This condition ensures that the pressure increases with density, preventing
the system from separating into regions of different densities.

The (isothermal) nuclear incompressibility parameter is evaluated
at saturation density,
\[
K_0 = 9 \left. \frac{\partial P}{\partial \rho_B} \right|_{\rho_0}.
\]
If  $K_0<0$ the system becomes mechanically unstable. This corresponds to a region
of negative incompressibility and signals the onset of a spinodal
instability, where uniform nuclear matter separates into phases of
different densities \cite{ref2}.

Since the effective mass $M^{*}$ enters directly into the energy density and single-particle spectrum, multiple self-consistent solutions for $M^{*}$ at a given $k_F$ correspond to competing branches of the equation of state. In the bifurcation diagrams, the coexistence of periodic or chaotic attractors therefore reflects the existence of multiple thermodynamically admissible (or metastable) solutions for the same density.

When a crisis occurs and a chaotic attractor is suddenly destroyed, the system loses one of these branches of self-consistent solutions \cite{crisis1,crisis2,crisis3}. From the thermodynamic viewpoint, this can correspond to the disappearance of a metastable minimum in the energy density, i.e., a restructuring of the equation of state. Such an event may coincide with a change in the sign or magnitude of $\partial P / \partial \rho$, indicating a modification of the compressibility and possibly the boundary of a spinodal instability region.

In this sense, the crisis can be interpreted as the dynamical signature of a macroscopic rearrangement of the scalar mean field that impacts the stiffness of nuclear matter. In extreme cases at high density, the strong reduction of $M^{*}$ accompanying these nonlinear effects may also be viewed as a precursor of partial chiral symmetry restoration or as an indication of an instability of the hadronic phase.

Since Fig.\ref{fig2} displays a very sensitive behavior of the routes to chaos of the birurcation diagrams acoording to the control parameters, we constructed some curves in order to identify where the first bifucartion occurs, where the attractor vanishes and waht is the value of $M_n$ when the first bifurcation occurs.

\begin{figure*}[htb]
\begin{center}
\centerline{\includegraphics[width=18cm,height=13cm]{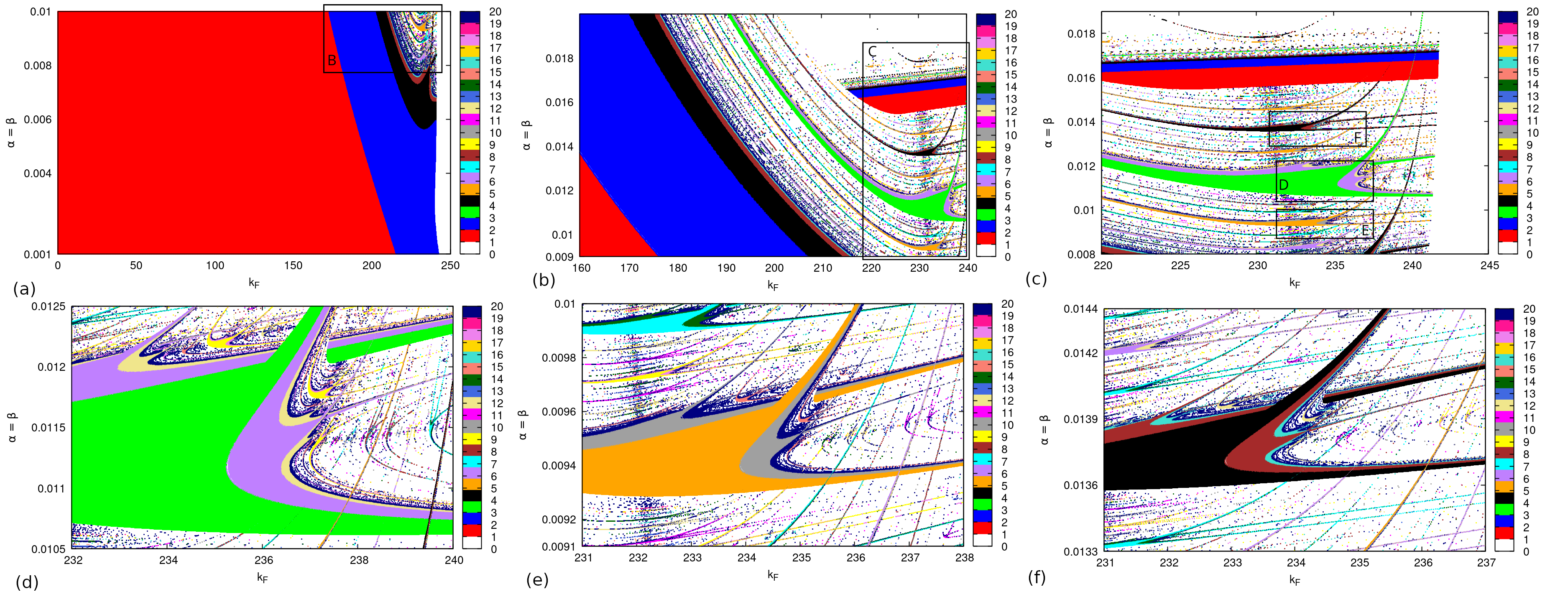}}
\end{center}
\caption{Color online: {\it High resolution discretization of the parameter space $(k_F, \alpha = \beta)$. In (a) there is an overview of the parameter space illustrating the transition from large scale periodic regimes (period 1 in red and period 2 in blue) into higher periods and chaotic domains. In (b)--(f) Successive magnifications of the regions indicated by boxes, revealing a complex hierarchy of shrimp shaped domains. These structures consist of self-similar, isoperiodic stable windows embedded within chaotic regions. The color scale represents the temporal period of the orbits, where the ``body'' of each shrimp (e.g., period 6 in (d) or period 8 in (e)) undergoes a period doubling bifurcation cascade as it approaches the chaotic boundaries. The observed fractal repetition of these domains across various scales suggests a universal organizational feature of the system.}}
\label{fig5}
\end{figure*}

We display in Fig.\ref{fig4} the characteristic critical scaling thresholds of the bifucartion diagrams, as function of the control parameters. In Figure \ref{fig4} the bullets represents when occurs the first biffurcation of each pair $(k_F,\alpha=\beta)$ and the squares indicates when the attractor vanishes. The region comprehended between then, is the region of interest, where the doubling period route to chaos coexists for the combination of the control parameters. Also, displayed in diamonds in Fig.\ref{fig4}, is the value of $M_n$, when the first bifurcation occurs (it is the same value of the axis in $k_F$). One can see that, as $\alpha=\beta$ diminishes, the doubling of $M_n$ also gets smaller. If we extrapolate the values of $\alpha=\beta$ for even small values, over then $10^{-4}$, one could interplay a connection between the curves, which indicates that there would not be any chaos, or attractors for such values, only a constant value of $M_n$.

\subsection{Shrimp-Shape domains}

Bifurcations serve as transitions mechanism in nonlinear systems, where some critical threshold in the control parameters may radically transform the nature of the solutions of the system \cite{bifurca0,bifurca1,bifurca2}. In the parameter space, these transitions are not random; they give rise to stable periodic domains known as "shrimps." These structures act as islands of order embedded within chaotic regions, defined by specific bifurcation sequences that organize the temporal evolution of the system between regular and irregular regimes. The shape of these regions is defined by self-similarity, a fractal property where shrimp-shaped structures repeat across different scales as nonlinearity increases. This repetition reveals a deep geometric order: the system replicates its stability patterns infinitely, kepping the same structure. Such fractal behavior in the parameter space shows that the transition between order and chaos follows an organized, recurring hierarchy, regardless of the scale of observation.

In our simulations, we identified the presence of shrimp-shaped structures in the parameter space, indicating regions of stable periodic behavior embedded within chaotic domains, according an initial range of the control parameters set up in Fig.\ref{fig4}. These structures are displayed in Fig.\ref{fig5}, where the color palette scale is defined as function of the period oh the orbits. Each color represents a period, and white regions are defined as chaos (period higher then $100$.).

Figure \ref{fig5}(a) presents a broad view of the parameter space $(k_F, \alpha = \beta)$, dominated by large low-period regions. The primary red domain represents a period-1 regime, which transitions into a blue period-2 region as the nonlinearity parameter $k_F$ increases. Figure \ref{fig5}(b) is a zoom into the region marked ``B'' in Fig. \ref{fig5}(a) revealing the emergence of higher-order periodic structures. The transition from the large period-2 (blue) and period-4 (black) zones begins to show the complex interlocking of shrimp-shaped domains as the control parameters and the nonlinearity of the system interact.

Figure \ref{fig5}(c) displays the emergence of the shrimps, focusing on the area marked 'C' in Fig.\ref{fig5}(b), showing a rich mosaic of periodic windows embedded in chaos. Multiple isoperiodic stable structures are visible, including various smaller colored shrimps representing higher periods. Finally, Figs.\ref{fig5}(d,e,f) shows zoom-in windows of different periods, they are respectively amplifications of the areas marked as 'D', 'E' and 'F' of Fig.\ref{fig5}(c). One can see in all Figs.\ref{fig5}(d,e,f), cascades of period doubling, self similarity, and multiple period shrimps nested within chaotic regions.

The emergence of shrimps suggests the existence of robust periodic windows that persist under small variations of the control parameters, highlighting the intricate organization of stability within the broader chaotic landscape.

\section{Discussion, Final Remarks and Conclusions}
\label{sec4}

In this work, we investigated the nonlinear structure and stability properties of the self-consistent effective-mass equation of the nonlinear Walecka model through its associated fixed-point iterative map. The emergence of bifurcations, periodic windows, chaotic regimes, and crisis phenomena reveals that the scalar mean-field sector may exhibit a rich mathematical structure beyond the conventional monotonic behavior typically assumed in relativistic mean-field approaches. Although these features arise at the level of the iterative map defining $M^{*}$, their physical relevance must ultimately be assessed through their thermodynamic consequences.

It is important to emphasize that the nonlinear analysis developed in the present work should not be interpreted as evidence for experimentally established chaotic dynamics in neutron-star matter. At present, there is no direct observational indication that dense hadronic matter undergoes real-time chaotic evolution or crisis phenomena in astrophysical environments. The objective of this study is instead of a different nature: namely, to investigate the mathematical structure and stability properties of the nonlinear self-consistent equations generated by the nonlinear Walecka model.

In relativistic mean-field theories, the effective nucleon mass is obtained through nonlinear self-consistency conditions whose solutions determine the thermodynamic properties of the equation of state. Such nonlinear equations may admit multiple competing branches, metastable solutions, or loss of fixed-point stability under parameter variation. The existence of these structures is intrinsically relevant from the theoretical viewpoint because they directly affect the organization of the admissible mean-field solutions and the corresponding thermodynamic behavior of dense matter.

The iterative map analyzed in this work therefore constitutes a mathematical probe of the nonlinear solution manifold of the model. Within this interpretation, bifurcations indicate qualitative reorganizations of the self-consistent scalar mean field, while crisis-like transitions correspond to abrupt changes in the stability structure of the iterative solutions. These phenomena are not interpreted here as direct microscopic dynamical evolution of neutron-star matter, but rather as signatures of the complexity of the underlying nonlinear mean-field equations.

Nevertheless, although the present work does not claim observational evidence for chaos in neutron stars, nonlinear modifications of the scalar mean field may still have indirect thermodynamic consequences through their impact on the equation of state, compressibility, and stability of dense nuclear matter. Therefore, the investigation of the nonlinear structure of the self-consistent solutions remains physically relevant, particularly in the context of strongly interacting matter at supranuclear densities.

A fundamental constraint is provided by the existence of neutron stars with masses close to $2\,M_\odot$ \cite{Demorest2010,Antoniadis2013,Cromartie2020}. These observations require the high-density EoS to remain sufficiently stiff to sustain such large gravitational masses. Any nonlinear mechanism --- including crisis-induced restructuring of the scalar mean field --- that significantly softens the pressure at several times nuclear saturation density must therefore remain compatible with the empirical bound $M_{\mathrm{max}} \gtrsim 2\,M_\odot$. This condition restricts the allowed parameter space of the nonlinear couplings.

Additional constraints arise from measurements of neutron-star radii and tidal deformabilities obtained through NICER observations and gravitational-wave detections \cite{Abbott2018Tidal,Miller2021,Riley2021}. The tidal deformability parameter $\Lambda$ is highly sensitive to the pressure in the density range $\sim 1$--$3\,\rho_0$. Since nonlinear scalar-field dynamics directly influence the compressibility,
$
{\partial P}/{\partial \rho_B},
$
a crisis or bifurcation in the effective-mass sector would translate macroscopically into localized softening or stiffening of the EoS. Such modifications could produce observable changes in the mass–radius relation or in the $\Lambda(M)$ curve. Current observational bounds therefore indirectly limit the magnitude of nonlinear dynamical effects in dense matter.

Complementary information is provided by terrestrial nuclear experiments. Measurements of the giant monopole resonance (GMR) in finite nuclei determine the incompressibility modulus at saturation density, $K_0 \approx 220$--$260$ MeV \cite{Youngblood1999,Colo2004}. Because nonlinear self-interactions of the scalar field modify the curvature of the energy density and hence the incompressibility,
$
K ,
$
the parameter region compatible with chaotic or bifurcation behavior must also reproduce empirical values of $K_0$. This requirement significantly constrains the nonlinear sector near saturation.

In summary, while the dynamical analysis reveals that the nonlinear Walecka model admits a complex bifurcation structure in its self-consistent solutions, any physically viable realization of such behavior must satisfy stringent observational and experimental constraints. The existence of two-solar-mass neutron stars, limits on tidal deformability and radii, and empirical determinations of nuclear incompressibility collectively bound the thermodynamic impact of nonlinear scalar-field dynamics. Although no direct signature of chaos has been observed, astrophysical data provide meaningful indirect tests of the underlying equation of state and thus of the nonlinear regimes explored here.

Future work may extend this analysis to beta-equilibrated neutron-star matter and investigate whether specific nonlinear scenarios generate distinctive features in mass–radius or tidal-deformability relations that could be probed by upcoming multimessenger observations.

\acknowledgments
LAB and ALPL acknowledges FAPESP, CNPq and CAPES for financial support.  This research was supported by resources supplied by the Center for Scientific Computing (NCC/GridUNESP) of the S\~ao Paulo State University (UNESP).

\end{document}